\documentclass[12pt,a4paper]{article}
\usepackage{amsmath,amssymb,amsthm,epsfig}
\usepackage{epsfig}
\renewcommand{\theequation}{\thesection.\arabic{equation}}
\newlength{\extraspace}
\newlength{\extraspaces}
\newcommand{\be}{\begin{equation}
\addtolength{\abovedisplayskip}{\extraspaces}
\addtolength{\belowdisplayskip}{\extraspaces}
\addtolength{\abovedisplayshortskip}{\extraspace}
\addtolength{\belowdisplayshortskip}{\extraspace}}
\newcommand{\ee}{\end{equation}}

\newcommand{\ba}{\begin{eqnarray}
\addtolength{\abovedisplayskip}{\extraspaces}
\addtolength{\belowdisplayskip}{\extraspaces}
\addtolength{\abovedisplayshortskip}{\extraspace}
\addtolength{\belowdisplayshortskip}{\extraspace}}
\newcommand{\ea}{\end{eqnarray}}  

\newcommand{\bas}{\begin{eqnarray*} 
\addtolength{\abovedisplayskip}{\extraspaces}
\addtolength{\belowdisplayskip}{\extraspaces}
\addtolength{\abovedisplayshortskip}{\extraspace}
\addtolength{\belowdisplayshortskip}{\extraspace}}
\newcommand{\eas}{\end{eqnarray*}}

\newcounter{subequation}[equation]
\makeatletter

\expandafter\let\expandafter
\reset@font\csname reset@font\endcsname
\def\subeqnarray{\arraycolsep1pt
    \def\@eqnnum\stepcounter##1{\stepcounter{subequation}%
        {\reset@font\rm(\theequation\alph{subequation})}}
\jot5mm     \eqnarray}

\def\subarray{\arraycolsep1pt
    \def\@eqnnum\stepcounter##1{\stepcounter{subequation}%
        {\reset@font\rm(\alph{subequation})}}
\jot5mm     \eqnarray}

\makeatother

\newcommand{\newsection}[1]{
\vspace{15mm}
\pagebreak[3]
\addtocounter{section}{1}
\setcounter{equation}{0}
\setcounter{subsection}{0}

\begin{flushleft}
{\large\bf \thesection. #1}
\end{flushleft}
\nopagebreak
\medskip
\nopagebreak}

\newcommand{\Z}{\mathbb{Z}}
\newcommand{\R}{\mathbb{R}}

\renewcommand{\H}{\mathbb{H}}

\newcommand{\bra}{\langle} 
\newcommand{\ket}{\rangle} 
\newcommand{\ra}{\rightarrow}

\newcommand{\is}{ &\! =\! & }
\newcommand{\nonum}{\nonumber \\[1.5mm]}
\newcommand{\sspace}{\makebox[1cm]{ }}    
\newcommand{\bspace}{\makebox[2cm]{ }}

\newcommand{\Tr}{{\rm Tr}}

\renewcommand{\th}{{\theta}} 
\newcommand{\eps}{\epsilon}
\newcommand{\lb}{\lambda}
\newcommand{\om}{\omega}
\newcommand{\sh}{{\rm sh}}
\newcommand{\ch}{{\rm ch}}
\newcommand{\dd}{{\partial}}

\newcommand{\cA}{{\cal A}}

\newcommand{\cD}{{\cal D}}

\newcommand{\cN}{{\cal N}}
\newcommand{\cM}{{\cal M}}

\begin{document}


\renewcommand{\thefootnote}{\fnsymbol{footnote}}
\begin{flushright}
\end{flushright}
\mbox{}
\vspace{2mm}

\begin{center}
\mbox{{\Large \bf The large $N$ expansion}}\\[4mm]
{{\Large \bf  in hyperbolic sigma-models}}
\vspace{1.3cm}

{{\sc M.~Niedermaier\footnote{Membre du CNRS} and  E. Seiler}}
\\[4mm]
{\small\sl Laboratoire de Mathematiques et Physique Theorique}\\
{\small\sl CNRS/UMR 6083, Universit\'{e} de Tours}\\
{\small\sl Parc de Grandmont, 37200 Tours, France}
\\[3mm]
{\small\sl Max-Planck-Institut f\"{u}r Physik}\\
{\small\sl F\"ohringer Ring 6}\\
{\small\sl 80805 M\"unchen, Germany}
%
%
\end{center}
{\small
{\bf Abstract.} Invariant correlation functions for ${\rm SO}(1,N)$ 
hyperbolic sigma-models are investigated. The existence of a large $N$ 
asymptotic expansion is proven on finite lattices of dimension $d \geq 2$. 
The unique saddle point configuration is characterized by a negative gap 
vanishing at least like $1/V$ with the volume. Technical difficulties  
compared to the compact case are bypassed using horospherical coordinates 
and the matrix-tree theorem.} 



\newsection{Introduction}

Noncompact sigma-models differ in several non-manifest ways from compact 
ones. Among the differences is the fact that in a large $N$ expansion of 
the $d\geq 2$ dimensional lattice systems the dynamically generated gap is 
negative and vanishes in the limit of infinite lattice size 
\cite{DNS,NSW}. The termwise defined infinite volume limits of invariant 
correlation functions also do not show exponential clustering 
\cite{NSW,TDlimit}. The justification of the large $N$ expansion in the 
noncompact models likewise has to proceed differently as the dualization 
procedure familiar from the compact models is ill-defined.
 
The goal of this note is to present a solid justification of the $1/N$ 
expansion for ${\rm SO}(1,N)$ invariant nonlinear sigma-models on a finite 
lattice. The main result is a proof that the $1/N$ expansion of invariant 
correlation functions is asymptotic to all orders in any finite volume 
$V=L^d,\,d \geq 2$. The `dual' action used to generate the expansion 
arises by performing Gaussian integrals in horospherical coordinates, 
thereby reducing the number of dynamical degrees of freedom per site from 
$N$ to $1$. This dual action served in \cite{NSW} to relate the large 
$N$ coefficients of the ${\rm SO}(1,N)$ model to those of its compact 
${\rm SO}(N+1)$ invariant counterpart. Here we show that the saddle 
point previously used in \cite{NSW,TDlimit} is in fact a {\it global 
minimum} of the dual action and the {\it only} critical point in 
the domain of integration. The strategy invokes a convexity argument 
based on Kirchhoff's matrix-tree theorem.


\newsection{Invariant correlators via horospherical coordinates} 

We consider the ${\rm SO}(1,N)$ 
hyperbolic sigma-models with standard lattice action, defined on 
a hypercubic lattice $\Lambda \subset \Z^d$ of volume 
$V = |\Lambda| = L^d$. The dynamical variables (``spins'') 
will be denoted by $n_x^a$, $x \in \Lambda$, $a =0, \ldots, N$,  
and periodic boundary conditions are assumed throughout 
$n_{x + L\hat{\mu}} = n_x$. The target manifold is  
the upper half of the two-sheeted $N$-dimensional hyperboloid, i.e. 
\ba
\H_N &=& \{ n \in \R^{1,N}\,|\, n \cdot n =1,\, n^0 >0\}\,, 
\nonum
a \cdot b &=& a^c \eta_{cd} b^d = a^0 b^0 - a^1 b^1 - \ldots - a^N b^N = 
a^0 b^0 - \vec{a} \cdot \vec{b}\,. 
\label{def1}
\end{eqnarray}
As indicated we shall also use the notation 
$\vec{a} =(a^1, \ldots, a^N)$ for vectors in $\R^{N}$.
The isometry group of $\H_N$ is ${\rm SO}_0(1,N)$. 

In terms of the hyperbolic spins the lattice action reads 
\be 
S =  \beta \sum_{x,\mu} (n_x \cdot n_{x+\hat{\mu}} -1) = 
- \frac{\beta}{2} \sum_x n_x \cdot (\Delta n)_x \geq 0\,,
\label{def2}
\end{equation}
where $\beta>0$ and $\Delta_{xy} = - \sum_{\mu} [2 \delta_{x,y} - 
\delta_{x,y + \hat{\mu}} - \delta_{x, y - \hat{\mu}}]$, as usual. 
We write 
\be
d\Omega(n) = 2 d^{N+1} n\,\delta(n\cdot n -1) \theta(n^0)\,,
\label{def3}
\end{equation}
for the invariant measure on $\H_N$. 

The goal in the following is to describe the invariant correlation functions 
$\bra n_{x_1} \cdot n_{y_1} \ldots n_{x_r} \cdot n_{y_r}\ket$
for the lattice statistical field theory with dynamical 
variables $n_x, \,x \in \Lambda$, and action (\ref{def2}). 
This is conveniently done in terms of a generating functional. 
Since the invariance group 
${\rm SO}_0(1,N)$ of the action (\ref{def2}) has infinite volume, 
the usual generating functional for connected invariant correlation 
functions is ill defined. A technically convenient way to gauge fix is 
to hold one spin, say $n_{x_0},\, x_0 \in \Lambda$, fixed. Then 
no Faddeev-Popov determinant arises and only the complications 
coming from the superficial lack of translation invariance have to 
be dealt with. We therefore consider the following generating 
functional 
\be
\exp W[H] = \cN \!\int\! \prod_{x} d\Omega(n_x) 
\delta(n_{x_0},n^{\uparrow})\,
\exp\Big\{\!
- S + \frac{1}{2} \sum_{x,y} H_{xy}( n_x \cdot n_y -1) \Big\}\,.
\label{Wdef}
\end{equation}
Here $\delta(n,n')$ is the invariant point measure on $H^N$,
$n^{\uparrow} = (1,0,\ldots,0)$, and sources $H_{xy} = H_{yx} <0$, $H_{xx} =0$,
give damping exponentials. The normalization $\cN = \cN[H]$ is 
chosen such that $\exp W[0] = 1$.  Connected $2r$ point functions are 
defined by 
\ba 
&& W[H] = \sum_{r \geq 1} \frac{1}{r! \, 2^r}\, 
W_r(x_1,y_1;\ldots;x_r, y_r) \, H_{x_1 y_1} \ldots 
H_{x_r y_r} \,,
\nonum
&& W_r(x_1,y_1;\ldots;x_r, y_r) := h_{x_1 y_1} \ldots h_{x_r y_r}
W[H]\Big|_{H=0}\,,\quad h_{xy} := \frac{\delta}{\delta H_{xy}}\,.
\label{def4}
\end{eqnarray}
In particular $W_1(x,y) = \bra n_x \cdot n_y\ket -1$, 
$W_2(x_1,y_1;x_2, y_2) := \bra n_{x_1} \cdot n_{y_1}  n_{x_2} \cdot
n_{y_2}\ket - \bra n_{x_1} \cdot n_{y_1} \ket \bra n_{x_2} \cdot
n_{y_2}\ket$, where $\bra \;\;\ket$ are the functional averages 
with respect to $\cN^{-1} e^{-S}$. Note that 
$W_r(\ldots; x,x; \ldots) =0$.

In the above we tacitly assumed that $W[H]$ and the correlation functions
computed from it do not depend on the site $x_0$ of the frozen
spin and are translation invariant. If we momentarily indicate the
dependence on the site as $W_{x_0}$ one has trivially 
\be 
W_{x_0}[\tau_a H] = W_{x_0 + a}[H]\,,\quad 
(\tau_a H)_{xy} = H_{x+a,y+a}\,.
\end{equation}
Thus, if $W_{x_0}$ is independent of $x_0$ it is also translation
invariant. The Bolzmann factor $f$ in (\ref{Wdef}) can be viewed as 
a function on the group via $F(g_0, \ldots, g_n) = f(g_0 n^{\uparrow}, 
\ldots , g_n n^{\uparrow})$, where we picked some ordering of the 
sites $x_i, \, i =0,\,1, \ldots , s :=V \!-\!1$, and identified $n_{x_i}$ with
$g_i n^{\uparrow}$. Then $W_{x_i}$ is of the form  
\ba
&& \int \!\prod_j d \Omega(n_j)\, \delta(n_i, n^{\uparrow}) 
f(n_1, \ldots, n_s)  = {\rm const} \,U_i\,,
\nonum
&&  U_i :=  \int\! \prod_{j \neq i} dg_j \, F(g_1, \ldots, g_{i-1}, e,
g_{i+1} , \ldots, g_s) \,.
\end{eqnarray} 
Using the invariance of $F$ under $g_i \mapsto h^{-1} g_i$ and the
unimodulariy of the measure $dg$ one verifies: $U_i = U_0$ for all 
$i$.     

The hyperboloid $\H_N$ admits an alternative parameterization in 
terms of so-called horospherial coordinates. These arise 
naturally from the Iwasawa decomposition of ${\rm SO}_0(1,N)$. 
Here it suffices to note the relation to the hyperbolic spins
\be 
n^0 = \ch \th + \frac{1}{2} t^2 e^{-\th}\,,
\quad 
n^1 = \sh \th + \frac{1}{2} t^2 e^{-\th}\,,
\quad n^i = e^{-\th} t_{i-1},\;\;i = 2,\ldots, N\,,
\label{def5}
\end{equation} 
and that $\H_N \ni n \mapsto (\th, t_1,\ldots ,t_{N-1}) \in \R^N$  
is a bijection. It is convenient to write $\vec{t} = (t_1, \ldots, t_N)$ 
and $\vec{t}\cdot \vec{t}' = t_1 t_1' + \ldots t_N t_N'$. 
For the dot product of two spins $n_x,\,n_y \in \H_N$ this gives 
\be 
n_x \cdot n_y = \ch(\th_x - \th_y) + \frac{1}{2} (\vec{t}_x - 
\vec{t}_y)^2 
e^{-\th_x - \th_y}\,,
\label{def6}
\end{equation} 
and for the measure (\ref{def3}) 
\be 
d\Omega(n) = e^{-\th(N-1)} d\th \,dt_1\ldots dt_{N-1} = 
e^{-\th(N-1)} d\th \,d \vec{t}\,. 
\label{def7}
\end{equation}
The key advantage of horospherical coordinates is manifest from 
(\ref{def6}), (\ref{def7}): for a quadratic action of the form 
(\ref{def2}) the integrations over the $\vec{t}$ variables are 
Gaussian and can be performed without approximations.  
The result is summarized in the 

PROPOSITION 2.1. {\it The generating functional (\ref{Wdef}) can be 
rewritten as  
\ba
\label{Wa1} 
&& \exp W[H]= \exp\Big\{ - \frac{1}{2} \sum_{x,y} H_{xy} \Big\} 
\,\cN \!\int_{\cD(H)} \prod_{x \neq x_0} da_x 
\nonum 
&& \quad \times 
\exp\Big\{ - \frac{N+1}{2} \Tr \ln \widehat{A} - \frac{\beta}{2} 
\sum_{x \neq x_0} a_x + \frac{\beta}{2}  (\widetilde{A}^{-1})_{x_0x_0}^{-1}
\Big\}\,.
\end{eqnarray} 
Here 
\be 
\label{Wa2}
A_{xy} = - \Delta_{xy} + \frac{1}{\beta} H_{xy} + \delta_{xy} a_x 
= \widetilde{A}_{xy} + a_{x_0} \delta_{x x_0} \delta_{xy}\,, 
\end{equation}
and $\cD(H)$ is an open set given by
\be
\label{Wa3}
\cD(H) = \{ \;a \in (2d, \infty]^{V-1}\;|\;\; \widehat{A} \;
\mbox{positive definite}\; \}\,.
\end{equation}}
\vspace{-4mm}

{\it Remarks.} (i) Compared to (\ref{Wdef}) the number of dynamical 
variables per site has been reduced from $N$ to $1$. 

(ii) The $H$-dependence of the domain $\cD(H)$ will produce extra
contributions in the variations with respect to $H$ defining the 
multipoint functions. Their direct computation is cumbersome but
their form can be inferred by first varying (\ref{trans6}) and 
then changing variables as before. For example 
\ba 
&& \frac{\dd}{\dd H_{xy}} \exp\Big\{ W[H] + \frac{1}{2} 
\sum_{xy} H_{xy} \Big\} 
\\
&& \quad = \Big\langle - \frac{N\!-\!1}{2 \beta} 
\Big[ 2 (\widehat{A}^{-1})_{xy} 
-  (\widehat{A}^{-1})_{xx} \frac{r_y}{r_x} - (\widehat{A}^{-1})_{yy}
\frac{r_x}{r_y} \Big] + \frac{1}{2} \Big( \frac{r_x}{r_y} +
\frac{r_y}{r_x} \Big) \Big\rangle\,,
\nonumber
\end{eqnarray}
where $r_x = r_x(a,H)$ is given by (\ref{trans9}) below. This is to be compared
with  the right hand side arising by varying (\ref{Wa1}), i.e.~$\bra - \lambda 
(\widehat A^{-1})_{xy} +r_xr_y + {\rm boundary}\;{\rm  terms} \ket$. 
As we shall see below in a large $N$ expansion the boundary terms do 
not contribute and (\ref{Wa1}) is a convenient starting point for such an
expansion. 

Underlying the Proposition is a nonlocal change of variables for which 
we prepare the 

LEMMA 2.2. {\it (a) Defining $a_{x_0}$ via (\ref{Wa2}) the condition 
$\det A =0$ is equivalent to 
\be 
a_{x_0} = - \frac{\det \widetilde{A}}{\det \widehat{A}} = 
- \frac{1}{(\widetilde{A}^{-1})_{x_0x_0}} \,,
\label{trans10}
\end{equation}
thereby determining $a_{x_0}$ as a function of $a_x, \,x \neq x_0$. 
\\
(b) The map 
\ba
&&\chi: \R^{V-1} \ra \cD(H)\,,\bspace \th_x \mapsto a_x \,\quad x \neq x_0 \,,
\nonum
&& a_x := \frac{1}{r_x} [ (\Delta - \beta^{-1} H) r]_x \,,\quad \;\;
\;\,r_x = e^{-\th_x}\,,\;\;x \neq x_0\,,\; \;r_{x_0} =1\,,
\label{trans7} 
\end{eqnarray}
with $\cD(H)$ as in (\ref{Wa3}) is a diffeomorphism.}

{\it Proof.} (a) Laplace expansion with 
respect to the $x_0$-th row gives $\det A = (2d + a_{x_0}) \det \widehat{A} + 
R$, where $R$ is the contribution from the columns $x \neq x_0$. 
Similarly $\det \widetilde{A} = 2d \det \widehat{A} + R$, with the same 
$R$. Eliminating $R$ gives $\det A - \det \widetilde{A} = 
a_{x_0} \det \widehat{A}$, and using $\det \widetilde{A} = \det \widehat{A}
/(\widetilde{A}^{-1})_{x_0x_0}$ one finds (\ref{trans10}).

(b) We define 
\ba
\label{cAdef}
\cA_{xy} &:=& \cM_{xy}  + \frac{1}{\beta} H_{xy} - \frac{1}{\beta} 
\delta_{xy} 
\sum_z e^{\th_x - \th_z} H_{xz} \,,
\nonum
\cM_{xy} &:=& - \Delta_{xy} + \delta_{xy} 
\sum_{\mu}( e^{\th_x - \th_{x+\hat{\mu}}} + e^{\th_x - \th_{x-\hat{\mu}}}-2)\,, 
\end{eqnarray} 
and write $\widehat{\cA}$ for the matrix obtained from $\cA$ 
by deleting the $x_0$-th row and column. Then $\cA$ has a null eigenvector    
\be 
\sum_y \cA_{xy} \,e^{-\th_y} =0\,,\quad \det \cA =0\,,
\label{trans5} 
\end{equation} 
but $\widehat{\cA}$ has maximal rank and is is positive definite. 
To see the latter it suffices to note that 
\be 
(r \vec{t}, \cA\, r \vec{t}\,)= 
\sum_{\bra x y \ket} (\vec t_x-\vec t_y)^2 r_x r_y - 
\frac{1}{2\beta} \sum_{x,y} (\vec t_x-\vec t_y)^2 H_{xy}
r_x r_y \geq 0\,, 
\label{Aform}
\end{equation}
is non-negative for $H_{xy} \leq 0$ and vanishes if and only if 
all $\vec t_x$ are equal. 

Further, by (a) 
\be 
\cA\circ \chi= A\,,
\label{chiA}
\end{equation}
provided $a_{x_0}$ is determined according to (\ref{trans10}). 
Since 
\be 
\frac{\dd a_x}{\dd \th_y} = e^{\th_x} \cA_{xy} e^{-\th_y} \,,
\quad 
\det\Big( \frac{\dd a_x}{\dd \th_y} \Big)_{x,y \neq x_0} = 
\det \widehat{\cA} >0\,,
\label{trans8}
\end{equation}
the  change of variables (\ref{trans7}) is locally invertible. 
Global invertibility is best seen from the inversion formula
\be 
\label{trans9}
r_x(a,H) = - \sum_{y \neq x_0} (\widehat{A}^{-1})_{xy} A_{yx_0}  
= \frac{(\widetilde{A}^{-1})_{x x_0}}{(\widetilde{A}^{-1})_{x_0 x_0}}\,, 
\quad x \neq x_0\,.
\end{equation}
The first equation follows from (\ref{trans5}), i.e.~$\sum_{y \neq x_0} 
\widehat{A}_{x y} r_y = - A_{x x_0} r_{x_0}$, assuming $r_{x_0} =1$. 
To derive the second expression in (\ref{trans9}) we 
extend the relation $A_{xy} = - \Delta_{xy} + \beta^{-1} H_{xy} +
\delta_{xy} a_x$, to $x,y= x_0$, and momentarily choose $a_{x_0}$ not as 
in (\ref{trans10}), but such that $\det A \neq 0$. Then (\ref{aux2}) below 
is applicable and gives $r_x = (A^{-1})_{x x_0}/(A^{-1})_{x_0x_0}$. 
On the other hand $\widetilde{A}_{xy} := A_{xy} - 
a_{x_0} \delta_{x x_0} \delta_{xy}$ is manifestly 
independent of $a_{x_0}$ and is nondegenerate. By (\ref{aux7}) below 
$r_x$ equals $(\widetilde{A}^{-1})_{x x_0}/
(\widetilde{A}^{-1})_{x_0 x_0}$, where one is free to adjust 
 $a_{x_0}$ such that 
$\det A =0$, as required by (\ref{trans5}), (\ref{chiA}).  
One can also insert (\ref{trans9}) into (\ref{trans5}) and finds 
\be 
\sum_y A_{xy} r_y(a,H) = \delta_{x x_0} [a_{x_0} +
  (\widetilde{A}^{-1})_{x_0x_0}^{-1} ]\,,
\end{equation} 
consistent with (a).

So far $\cD(H)$ entered as the image of $\R^{V-1}$ under $\chi$. 
By definition of $r_x=\exp(-\theta_x)$ the domain $\cD(H)$ is 
characterized by the 
condition $r_x(a,H) = (\widetilde{A}^{-1})_{x x_0}/ (\widetilde{A}^{-1})_{x_0 
x_0} >0$. We verify that this is also equivalent to the positive 
definiteness of the matrix $\widehat{A}$: First assume that all $r_x>0$. 
Then by (\ref{Aform}), (\ref{chiA}) $\widehat A$ is positive definite. 
Conversely, assume that $\widehat A$ is positive definite, but that 
there is a $y_0$ such that $r_{y_0}< 0$. Remembering $r_{x_0}=1$, 
choose $\vec t_x=\vec s\neq \vec 0$ for all $x$ satisfying $r_x>0$ and 
$\vec t_x=\vec 0$ for all other $x$. Then
\be
(r \vec{t}, A\, r \vec{t})=
\sum_{\bra xy\ket: r_xr_y < 0} \vec{s}^{\;2} r_x r_y
-  \frac{1}{2\beta}\sum_{x,y: r_xr_y< 0} \vec{s}^{\;2} H_{xy} r_xr_y<0\ .
\end{equation}
This is a contradiction, so $a \in \cD(H)$ if and only if 
$\widehat{A}$ is positive definite. By the Hurwitz (or   
Sylvester) criterion this is equivalent to 
\be 
\cD(0)=
\{a \in R^{V -1} \,|\, \det A_k >0\,,\;\forall
k=1,\cdots, V-1 \,,\;\; A_k=(A_{x_i x_j})_{1\le i,j\le k}\}\,,
\end{equation} 
where we picked an arbitrary ordering of the lattice 
sites $x_0,x_1,\cdots, x_{V-1}$. For $k=1$ one gets in  
particular $a_x > -2d$ for all $x \in \Lambda$ 
(recall $H_{xx} =0$) and  (\ref{Wa3}) follows. $\Box$

Before turning to the proof of the Proposition we prepare some simple
auxiliary results. Let $A = (A_{xy})_{x,y \in \Lambda}$ be a symmetric
invertible matrix such that the matrix $\widehat{A}$ arising from 
$A$ by deleting its $x_0$-th row and column is positive definite. 
Then 
\begin{eqnarray}  
\label{aux1} 
&& \int \prod_x d\phi_x \delta(\phi_{x_0})\, 
\exp\Big\{ - \frac{1}{2} \sum_{x,y} \phi_x A_{xy} \phi_y 
+ \sum_x J_x \phi_x\Big\}   
\nonum
&& \quad = (2 \pi)^{\frac{V-1}{2}} (\det \widehat{A})^{-1/2} 
\exp \Big\{ \frac{1}{2} \sum_{x,y} J_x (\widehat{A}^{-1})_{xy} J_y
\Big\}\,, 
\end{eqnarray}
for a real field $\phi_x, \, x \in \Lambda$. The inverse of $\widehat{A}$ 
can be expressed in terms of the inverse of $A$ via 
\be 
(\widehat{A}^{-1})_{xy} = (A^{-1})_{xy} - 
\frac{(A^{-1})_{xx_0}(A^{-1})_{yx_0}}{(A^{-1})_{x_0x_0}}\,.
\label{aux2}
\end{equation}    
The determinant of $\widehat{A}$ is related to that of $A$ by 
\be 
\det A = \frac{\det \widehat{A}}{ (A^{-1})_{x_0x_0}}\,. 
\label{aux4}
\end{equation}

Often a term in the $x_0$-th matrix 
element on the diagonal of $A$ has to be split off according to 
$A_{xy} = \widetilde{A}_{xy} - c \delta_{xy} \delta_{x_0 x}$.
In this case the inverse of $A$ is related to the inverse of 
$\widetilde{A}$ by 
\be 
(A^{-1})_{xy} = (\widetilde{A}^{-1})_{xy} + 
\frac{c}{ 1 - c (\widetilde{A}^{-1})_{x_0x_0}} 
(\widetilde{A}^{-1})_{x x_0}(\widetilde{A}^{-1})_{y x_0}\,.
\label{aux6}
\end{equation} 
In particular $A_{x_0x_0} - (A^{-1})_{x_0x_0}^{-1} = \widetilde{A}_{x_0x_0} - 
(\widetilde{A}^{-1})_{x_0x_0}^{-1}$ and 
\be
\frac{1}{(A^{-1})_{x_0 x_0}} = -c + \frac{1}{(\widetilde{A}^{-1})_{x_0x_0}}\,,
\sspace \frac{(A^{-1})_{xx_0}}{(A^{-1})_{x_0x_0}} = 
\frac{(\widetilde{A}^{-1})_{xx_0}}{(\widetilde{A}^{-1})_{x_0x_0}}\,.
\label{aux7}
\end{equation} 
For the determinants one has 
\be 
\det A = \det \widetilde{A} - c \det \widehat{A}\,.
\label{aux7b}
\end{equation}

{\it Proof of the Proposition.} We rewrite the action as 
\ba
\label{trans1}
S \is \beta \sum_{x,\mu} \Big[ \ch(\th_x - \th_{x + \hat{\mu}}) 
+ \frac{1}{2} (\vec{t}_x - \vec{t}_{x + \hat{\mu}})^2 
e^{-\th_x - \th_{x + \hat{\mu}}} -1 \Big] 
\nonum
\is \beta \sum_{x,\mu} \ch(\th_x - \th_{x + \hat{\mu}}) + 
\frac{\beta}{2} \sum_{x,y} e^{-\th_x - \th_y} \cM_{xy} \vec{t}_x \cdot \vec{t}_y
- \beta d\, V \,,
\end{eqnarray} 
with $\cM$ as in (\ref{cAdef}). The source term in (\ref{Wdef}) can be 
rewritten similarly and using also (\ref{def7}) one finds in a first step
\ba 
\label{trans3}
&& \exp W[H] = \cN \int \prod_x e^{-(N-1)\th_x} d\th_x \, 
\;e^{(N-1) \th_{x_0}} \delta(\th_{x_0})
\nonum
&& \quad \times  
\exp\Big\{ - \beta \sum_{x,\mu} \ch(\th_x - \th_{x + \hat{\mu}}) + 
\frac{1}{2} \sum_{x,y} H_{xy} [\ch(\th_x - \th_y) -1]\Big\}   
\nonum 
&& \quad \times \int \prod_x  d\vec{t}_x \delta(\vec{t}_{x_0}) 
\exp\Big\{ - \frac{\beta}{2} \sum_{x,y} e^{-\th_x} \vec{t}_x 
\cdot \cA_{xy} \, e^{-\th_y} \vec{t_y} \Big\}\,,  
\end{eqnarray} 
with $\cA_{xy}$ as in (\ref{cAdef}). After the rescaling $\vec{t}_x \mapsto e^{\th_x} 
\vec{t}_x$ the Gaussians are of the form (\ref{aux1}) and one 
obtains 
\ba 
&& \exp W[H] = \cN \int \prod_x  d\th_x \delta(\th_{x_0})
\, \exp\Big\{ - \frac{N-1}{2} \Tr \ln \widehat{\cA} \Big\}
\nonum
&& \quad \times 
\exp\Big\{ - \beta \sum_{x,\mu} \ch(\th_x - \th_{x + \hat{\mu}}) + 
\frac{1}{2} \sum_{x,y} H_{xy} [\ch(\th_x - \th_y) -1]\Big\}\,,   
\label{trans6}
\end{eqnarray} 
with a redefined $\cN$. Next one observes that the integration 
variables $\th_x$ only occur through the combination (\ref{trans7}). 
Indeed, $\cA_{xy} = - \Delta_{xy} + \beta^{-1} H_{xy} + \delta_{xy} a_x(\th)$, 
$\sum_{x, \mu} \ch(\th_x - \th_{x + \hat{\mu}}) = d V + \frac{1}{2}\sum_x r_x^{-1} 
(\Delta r)_x$, and $\sum_{x,y} H_{xy}\ch(\th_x - \th_y) = 
\sum_{x,y} r_x^{-1} H_{xy} r_y$. This suggests to change variables 
in (\ref{trans6}) from $\th_x,\, x\neq x_0$ to $a_x,\,x \neq x_0$. 
The change of variables has been prepared in Lemma 2.2. Combining 
(\ref{trans6}), (\ref{chiA}), (\ref{trans8}), (\ref{trans10}) one 
arrives at (\ref{Wa1}). $\Box$

\newsection{Large $N$ expansion for $W[H]$} 

Connected invariant correlation functions are defined via the 
moments of $W[H]$. In a large $N$ expansion $\lambda := (N\!+\!1)/\beta$
is kept fixed and we write 
\be 
W_r \sim \frac{\lb^r}{(N \!+ \!1)^{r-1}}
\sum_{s \geq 0} \frac{1}{(N \! + \! 1)^s}\, W_r^{(s)} \,.
\label{N1}
\end{equation}
The algorithm to compute the $W_r^{(s)}$ is as follows \cite{NSW}:
Define 
\begin{subeqnarray}
\label{saddle0}
a_x \is \om_x + \frac{u_x}{\sqrt{N+1}} \,,\quad u_x \in \R\,,
\\
\om_x \is \om_- + \lb \delta_{x x_0} \,,
\\
\om_- &\;\;& \mbox{solution of} \;\;(\ref{N21}), (\ref{N22})\,,  
\end{subeqnarray}
and consider the Laplace expansion of (\ref{Wa1}) around 
(\ref{saddle0}b) where $\cD(H)$ has been replaced by $\R^{V-1}$.  
Our main result is:

THEOREM 3.1.
The correlation functions $W_r$ admit an asymptotic 
expansion of the form (\ref{N1}) whose coefficients
$W_r^{(s)}$  are determined by the above algorithm and 
are translation invariant.

{\it Remarks.} (i) Both the effective Gaussian measure 
and the vertices of the expansion depend on $x_0$ but the 
$W_r^{(s)}$ are translation invariant. Once asymptoticity of 
the expansion has been shown this follows for all
$r,s$ from the correspondence to the compact model shown 
in \cite{NSW}.

(ii) The core fact underlying the asymptoticity is that the `dual' action 
\be
S[a,H]= \frac{1}{2}\Tr\ln\widehat A+ 
\frac{1}{2\lambda}\sum_{x\neq x_0} a_x- \frac{1}{2\lambda} 
(\widetilde{A}^{-1})_{x_0x_0}^{-1}\,,
\label{N3}
\end{equation}  
with $A$ and $\widetilde{A}$ as in (\ref{Wa2}) has a {\it unique minimum} in 
the domain $\cD(H)$, given by (\ref{saddle0}b,c). This holds despite 
the unusual feature that the gap $\om_-$ is negative and so is for nonzero 
momentum the 1-loop polarization function. The latter fact ensures 
that  (\ref{saddle0}b,c) is at least a local minimum of $S[a,H]$,
as has been shown algebraically in \cite{NSW}. 

(iii) A heuristic derivation of the algorithm based on a 
dualization procedure was outlined in Appendix C of \cite{TDlimit} using 
\cite{DNS}, where also the approach to the $N\to\infty$ limit was checked 
numerically. Substituting $a_x = 2 i \lb \alpha_x$ in (\ref{N3}) 
gives an effective action that can formally be obtained by 
mimicking the dualization procedure in the compact model,
see Appendix C of \cite{TDlimit}. The flip $\alpha_x \mapsto - \alpha_x,\,
\lb \mapsto - \lb$, then relates it to the dual action of the 
compact model, see \cite{NSW} for the relation between both 
large $N$ expansions.  

{\it Proof.} We establish consecutively: (a) $S[a,0]$ has at most one 
extremum in $\cD(0)$, which if it exists must be a minimum,  
(b) existence of an extremum  in $\cD(0)$, and (c) the fact that the 
asymptotic expansion (\ref{N1}) is unaffected by the replacement 
of $\cD(H)$ with $\R^{V-1}$.   

(a) Since the $H$-dependent terms in $A_{xy}$ are $O(1/(N\!+\!1))$ it 
suffices to treat $H=0$. We consider  the preimage of $S[a,0]$ 
under $\chi$ and show that it is a strictly convex function on 
$\R^{V-1}$. Thus we set 
\be
F(\theta):= 2 S[a(\theta),0]= \Tr\ln\widehat\cM 
+\frac{1}{\lb} \sum_x e^{\th_x} (\Delta e^{-\th})_x\,,  
\end{equation}
where $a_x(\th) = e^{\th_x} (\Delta e^{-\th})_x$ and $\cM$ is as before. 
To establish strict convexity of $F$ it suffices to
show that both terms in $F$ are separately strictly convex.%
For the second term this is manifest: shifting $\th_x \mapsto 
\th_x + \eps_x$ the term quadratic in $\eps_x$ is nonnegative as 
$- \Delta$ is positive semi definite. 

To show convexity of $\Tr\ln\widehat\cM$ we define 
$W = (w_{xy})_{x,y \in \Lambda}$, by
\be
w_{xy} := e^{-\th_x} \cM_{xy} e^{-\th_y}\,,
\end{equation}
which obeys $\sum_y  w_{xy} =0$ and has matrix elements 
\be
w_{xy} = \left\{
\begin{array}{ll} 
- e^{-\th_x - \th_{x \pm \hat{\mu}}}\,,\quad & y = x \pm \hat{\mu}\,,
\\
e^{-\th_x} \sum_{\mu} ( e^{-\th_{x + \hat{\mu}}} +  e^{-\th_{x -\hat{\mu}}})\,,\quad & x =y \,,
\\
0  \quad & {\rm otherwise}\,. 
\end{array} \right. 
\end{equation} 
Trivially $\ln \det \widehat{\cM} = 2 \sum_x \th_x + \ln \det W$, so 
that strict convexity of $\ln \det W$ implies that of $F$. 
$W$ has the form that makes the so-called matrix-tree theorem (see
e.g.~\cite{Bollobas,BurSha,Abdel}) applicable. The matrix-tree theorem 
then entails 
\be
\det \widehat W =\sum_T w_T\,,
\label{tree1}
\end{equation}
where the sum runs over all spanning trees built from nearest neighbor 
pairs, i.e.~walks through the lattice $\Lambda$ visiting every point of 
$\Lambda$ once and
\be 
w_T=\prod _{(x,x\pm\hat\mu)\in T}w_{x,x\pm\hat\mu}\,.  
\label{tree2}
\end{equation}
The point of this representation is that it expresses $\det \widehat W$ 
as a sum of exponentials in the $\theta$ variables; the (strict) convexity 
of $\Tr\ln\widehat W$ follows from the well-known fact: if $Z(\th):= 
\sum_i c_i e^{a_i \cdot \th_i}$, $a_i,\,\th_i \in \R^n$, $c_i \geq 0$, then 
$\ln Z(\th)$ is convex.

(b) Here we proceed in two steps. In a first step we rewrite the 
stationarity conditions for $S[a,0]$ in a more transparent form. In a 
second step we present a solution for them in $\cD(0)$.  

For the first step we define the matrices $M,\,\widehat M,\, 
\widetilde M$ as $\cM,\,\widehat\cM,\,\widetilde \cM$  expressed in 
the coordinates $a_x$ and with the critical point parameters $\om_x$ of 
(\ref{saddle0}) inserted, i.e.
\be
M:=\chi\circ \cM\Big|_{a_x \ra \om_x} = -\Delta_{xy} + \delta_{xy} \om_x\,,
\end{equation}
and similarly for $\widehat M,\,\widetilde M$. 
Note that $\det M =0$ by Lemma 2.2 and (\ref{trans5}).
We are looking for a critical point of $S[\om,0]= \frac{1}{2}\ln \det 
\widehat M+ \frac{1}{2\lb} \sum_x \om_x$ under the condition $\det M=0$. 
Introducing a Lagrange multiplier $\mu$ for the latter we consider 
\be
\tilde{F}(\om,\mu) := \ln \det \widehat M + \frac{1}{\lb} \sum_x \om_x + 
\mu \det M\,.
\end{equation} 
The conditions for a critical point (`saddle point equations') of 
$\tilde{F}$ are: 
\begin{subeqnarray}
&&\det M=0\,,
\\
&&\mu=-\frac{1}{\lb \det \widehat M}\,,
\\
&&\lb \widehat{M}_{xx}^{\rm co} + \det \widehat{M} - M_{xx}^{\rm co}=0\,,
\quad x\neq x_0\,,
\label{saddle1}
\end{subeqnarray}
where we denoted the cofactor matrix of $M$, $\widehat{M}$ by 
$M^{\rm co}$, $\widehat{M}^{\rm co}$, respectively. 

The conditions (\ref{saddle1}) simplify when expressed in terms of 
\be
D^{-1}_{xy}:= M_{xy}-\lb \delta_{xx_0}\delta_{xy}\,.
\label{defd}
\end{equation}
Indeed, using (\ref{aux7b}) for the cofactors one finds 
\be 
(D^{-1})^{\rm co}_{xx} = \left\{
\begin{array}{ll} M_{x_0x_0}^{\rm co} = \det \widehat{M}\,,\quad & x = x_0\,,
\\[2mm]
- \lb \widehat{M}^{\rm co}_{xx} + M^{\rm co}_{xx} \,,\quad & x \neq x_0\,.
\end{array}
\right.
\end{equation}
By (\ref{saddle1}c) also the $x \neq x_0$ cofactors reduce to $\det \widehat{M}$. 
Using  (\ref{aux7b})  once more for $\det D = - \det M + \lb \det \widehat{M}$
one sees that the saddle point equations (\ref{saddle1}) are equivalent to
\be
-\lb D_{xx} = 1 \,,\quad \forall x \,,
\label{saddle4}
\end{equation}
where the $x= x_0$ equation implements (\ref{saddle1}a).

Also the conditions characterizing $\cD(0)$ can be expressed in terms 
of $D$. From the proof of Lemma 2.2 we know that $r_x(\om,0) = 
(\widetilde{M}^{-1})_{xx_0}/(\widetilde{M}^{-1})_{x_0x_0}>0$ characterizes 
$\cD(0)$. 
On the other hand writing $(D^{-1})_{xy} = \widetilde{M}_{xy} + (\om_{x_0} -\lb) 
\delta_{xy} \delta_{xx_0}$, and applying (\ref{aux6}) one has 
\be
D_{xy}=(\widetilde M^{-1})_{xy}+\frac{\lb-\om_{x_0}}{1-(\lb-\om_{x_0})
(\widetilde M^{-1})_{x_0x_0}}(\widetilde M^{-1})_{xx_0}(\widetilde
M^{-1})_{yx_0}\,.
\label{DMtilde}
\end{equation}
Taking into account that $(\widetilde{M}^{-1})_{x_0x_0} =-1/\om_{x_0}$ one 
arrives at the following characterization: 
\be 
\{\om_x,\,x \neq x_0\} \in \cD(0) \quad \mbox{if and only if} \quad 
- \lb D_{x x_0} >0\,.
\label{domD}
\end{equation}

In a second step we now search for a the solution of Eq.~(\ref{saddle4}) 
satisfying $- \lb D_{x x_0} >0$. Eq.~(\ref{saddle4}) is a system of $V\!-\!1$ 
algebraic equations for the $V\!-\!1$ critical point parameters $\om_x\,,x \neq x_0$,
and difficult to tackle analytically.  But the translation invariant form of 
the equation suggests the translation invariant ansatz
\be
D^{-1}_{xy}=-\Delta_{xy} + \om\delta_{xy}\,,
\end{equation}
i.e.
\be 
\om_x = \om +\lb\delta_{x x_0}\,.
\label{omansatz} 
\end{equation}
The saddle point equations (\ref{saddle4}) then reduce to a single almost 
conventional gap equation for $\om$  
\be 
D_{xx} = \frac{1}{V} \sum_p \frac{1}{E_p + \om} = -\frac{1}{\lb}\,,
\label{N21}
\end{equation}
where the sum is over all $p = \frac{2\pi}{L}(n_1,\ldots, n_d)$,
$n_i = 0,1,\ldots, L\!-\!1$, and $E_p := 2 d 
- 2 \sum_\mu \cos(p \cdot \hat{\mu})$. 
From (\ref{N21}) it is clear that all solutions $\om$ must 
be negative. As shown in \cite{NSW} there is a unique 
root $\om = \om_-(\lb,V)$ of (\ref{N21}) characterized by the 
following two equivalent conditions:   
\begin{subeqnarray} 
&& - \frac{4}{2 d +1} \sin^2 \frac{\pi}{L} < \om_- < 0\,,
\\
&& - \lb D_{xy}\Big|_{\om = \om_-} \geq 1\,,
\quad \mbox{for all}\;\; x,y \,.  
\label{N22}
\end{subeqnarray} 
Since $- \lb D_{xx_0}|_{\om = \om_-}> 1$ for this 
solution it lies in $\cD(0)$.

(c) This can be seen from the following simple fact about saddle point 
expansions: Let $f \in C^{\infty}(\R^n)$ be such that
$\exp (N f(x))$ is integrable for all $N$ and obeys  
\be 
\quad {\rm grad}f(0) = f(0) =0\,,\sspace 
f(x) < - \delta \;\;\mbox{for}\;\; |x| > \eps\,.
\end{equation} 
Then the integral has a saddle point expansion of the form 
\be 
\int \! dx \, \exp(N f(x)) \sim \sum_{n \geq 0} \frac{a_n}{N^n}\,,
\label{ii7}
\end{equation}
and the expansion coefficients are insensitive to 
changes of the integrand bounded away from the saddle point: 
If $q \in L^1(\R^n) \cap L^{\infty}(\R^n)$, with $q(x) = 1$ for 
$|x| < \eps$, then 
\be 
\int \! dx \, q(x) \,\exp(N f(x)) \sim \sum_{n \geq 0} \frac{a_n}{N^n}\,.
\label{ii8}
\end{equation}
This completes the proof of the theorem. 
$\Box$

{\it Remarks.} 
(i) Eq.~(\ref{saddle4}) can be viewed as the normalization condition,
$-\lb D_{xx} =1$, of the leading order two-point function. 
In fact \cite{NSW}
\be
\bra n_x \! \cdot \! n_y\ket\Big|_{N = \infty} 
= - \lb \widehat{D}_{xy} +
\frac{\widetilde{D}_{x x_0}\widetilde{D}_{y
x_0}}{\widetilde{D}_{x_0x_0}^2} = - \lb D_{xy}\,,
\label{N7}
\end{equation}
where $\widetilde{D} = \widetilde{M}^{-1}$, $\widehat{D} = \widehat{M}^{-1}$.   
The first equality is obtained by evaluating $W[H]$ to leading order in 
$1/(N+1)$, the second equality follows by using (\ref{aux2}) in 
(\ref{DMtilde}). 

(ii) The number of terms in (\ref{tree1}) is given by 
$\det(- \widehat{\Delta})$,
which is sizeable even for small lattices (but less than the naive 
$(V \!-\!1)!$ number of terms), e.g.~for $d=2$, $L=3$ there are 
11664 spanning trees.   

(iii) In making the ansatz (\ref{omansatz}) we 
took the consistency with $\om_{x_0} = -1/(\widetilde{M}^{-1})_{x_0x_0}$ 
(Eq. \ref{trans10}) for granted. Here $(\widetilde{M}^{-1})_{x_0x_0}$ is 
a ratio of polynomials (Toeplitz determinants) of degree $V\!-\!1$ in 
$\om$. Its direct computation is cumbersome but by assuming $\om_{x_0} = 
\om + \lb$ and eliminating $\lb$ via (\ref{N21}) one sees that 
\be 
\om_{x_0}(\om) = \om - \Big[ \frac{1}{V} 
\sum_p \frac{1}{E_p + \om} \Big]^{-1}\,,
\label{N19}
\end{equation}  
on the solutions of (\ref{N21}). Equivalently (\ref{N21}) is 
such that $M_{xy} = -\Delta_{xy} + \om \delta_{xy} + \lb \delta_{x y} 
\delta_{x x_0}$ has zero determinant.

(iv) We remark that the large volume asymptotics of $\om_-$ is given by 
\cite{NSW}  
\be 
- V \om_-(\lb, V) = \left\{ 
\begin{array}{ll} \displaystyle{\frac{4\pi}{\ln V}} 
\Big(1 + O(1/\ln V)\Big) \quad 
& d =2\,,
\\[4mm]
\Big(\dfrac{1}{\lb} + C_d \Big)^{-1} + O(V^{-\frac{d-2}{d}}) 
\quad & d\geq 3\,,   
\end{array}
\right.
\label{gap5}
\end{equation}
where $C_d = \int_0^{2\pi} \frac{d^d p}{(2\pi)^d}\,\frac{1}{E(p)}$. 
In particular the gap $\om_-(\lb,V)$ vanishes in the infinite 
volume limit, in sharp contrast to the compact model. 

(v) Convexity of a translation invariant effective action for 
the ${\rm SO}(1,2)$ model coupled to a symmetry breaking external 
field was shown in \cite{SZ} by a technique not readily 
transferrable to the situation here.

The theorem and its proof have a number of interesting corollaries. 

COROLLARY 3.2. {\it All solutions of Eqs.~(\ref{saddle4}) satisfying 
$-\lb D_{x x_0} >0$ are constant:~$\om_x = \om, \, x \neq x_0$.} 

By inspection of examples one sees that the inequalities are essential 
for the validity of the result: nonconstant solutions outside the domain 
$\cD(0)$ can easily be found. Since (\ref{saddle4}) is a system of 
$V\!-\!1$ algebraic equations for $V\!-\!1$ unknowns a direct proof of 
Corollary 3.2 seems difficult.

COROLLARY 3.3. {\it All solutions of (\ref{N21}) other than $\om_-$ do
not lie in $\cD(0)$. The solution $\om=\om_-$ lies in  $\cD(0)$
and thus implies the positive definiteness of $\widehat M|_{\om = 
\om_-}$.}

We recall from \cite{NSW} the form of the Hessian of the action 
(\ref{N3}) at the extremum $\om_x = \om_-$, $x \neq x_0$  
\be
S_2[u,H] =  - \frac{1}{4} \sum_{x,y\neq x_0} u_x u_y 
[D_-(x\!-\!y)^2 - \lb^2 D_-(x\!-\!x_0)^2 D_-(y\!-\!x_0)^2] + 
\frac{\lb}{2} \sum_{x,y} H_{xy} D_-(x\!-\!y)\,.
\label{p1}
\end{equation}
Here $D_-(x-y) := D_{xy}|_{\om = \om_-}$ and the variables $u_x$ are those 
of (\ref{saddle0}). 

One can show that all the matrix elements in square brackets in 
Eq.~(\ref{p1}) are negative. On account of the theorem we have

COROLLARY 3.4. \quad $S_2[u,0] \geq 0\,.$

More directly than here it has been shown in Appendix A of \cite{NSW} that 
\be 
S_2[u, H] \geq 0 \,,
\label{p2}
\end{equation}
for all $H_{xy} \leq 0$ and all $u$ configurations,

COROLLARY 3.5. {\it The minimum of $S[a,0]$ cannot lie at 
the boundary of $\cD(0)$. More generally one has 
\be 
S[a,H] \ra + \infty \quad \mbox{as}\quad a \ra \partial \cD(H)\,,
\label{cor4}
\end{equation} 
where $\partial D(H)$ is the boundary of $\cD(H)$.}

To show (\ref{cor4}) this it suffices to establish that $\det \widetilde{A} = 
2d \det \widehat{A} - R$, where $R$ is bounded from below by a ($a_x$
independent) positive constant $\# C$. Indeed, using 
$a_x \leq -2d$ for all $x$, it then follows 
\be 
S[a,H] \geq 
\frac{1}{2} \ln \det \widehat{A} + \frac{1}{2\lb} 
\frac{\# C}{\det \widehat{A}} - 2d V\,, 
\label{ii2}
\end{equation} 
and the positive second term dominates as $a$ approaches the boundary of 
$\cD(0)$ in (\ref{Wa3}b). Slightly more generally one has : 

If $A_{xy} = -\Delta_{xy} + a_x \delta_{xy} + \beta^{-1} H_{xy}$, with 
$H_{xx} =0, \;H_{xy} \leq 0$, is a positive semidefinite matrix 
on a hypercubic lattice of linear size $L$, and $R:= (2 d + a_{x_0}) 
\det \widehat{A} - \det A$, then 
\be 
R \geq \sum_{{\rm cycles}\;C} \prod_{\bra xy\ket \in C} (1 -\beta^{-1} H_{xy})
\geq \#\; \mbox{cycles on} \;\;\Lambda\,. 
\label{ii3}
\end{equation} 
Here $C$ is the set of cycles, i.e.~closed oriented paths which connect only 
nearest neighbors and which visit each lattice point exactly once.
On a torus of dimension $d$ the number of these cycles is at least
$2d$. We omit the proof. 


\newsection{Conclusions} 

Our result establishes the existence of a large N asymptotic 
expansion for hyperbolic sigma-models in finite volume and 
provides the rationale for the computational algorithm used 
in \cite{NSW,TDlimit}. It would be desirable also to have a proof 
that the expansion is uniform in the volume, which would then 
imply that the termwise thermodynamic limit yields the correct 
asymptotic expansion of the model in infinite volume. Kupiainen \cite{kupi} 
managed to show the corresponding result for the compact ${\rm O}(N)$ models 
for the region of high temperature (higher than the critical temperature 
of the limiting spherical model), but his proof relies in an essential 
way on features absent in the hyperbolic models: in the ${\rm O}(N)$ 
models the large $N$ saddle point has a mass gap and exponential decay 
as long as one is in the high temperature regime. As emphasized before, 
this is not the case in the non-compact models. Direct computation 
indicates nevertheless the existence of a termwise thermodynamic
limit \cite{TDlimit}. The structure of this termwise thermodynamic
limit does not suggest the existence of an interacting scaling 
limit in the invariant sector of the theory. An important 
open problem is to prove or disprove this ``triviality''.  
\bigskip

{\it Acknowledgment:} E.S. is grateful to D. Brydges for giving him free 
instruction about the matrix-tree theorem.


\begin{thebibliography}{99}
%
\bibitem{Abdel} A. Abdessalam, The Grassmannian-Berezin calculus and 
theorems of the matrix-tree type, Adv. Apl. Math. {\bf 33} (2004) 51.  
%
\bibitem{Bollobas} B.~Bollobas, {\it Modern Graph Theory}, Springer, 
2nd edition, 2002.   
%
\bibitem{BurSha} Y. Burman and B. Shapiro, Around matrix-tree theorem,
Math.~Res.~Lett. {\bf 13} (2006) 761.
%
\bibitem{DNS} A. Duncan, M. Niedermaier, and E. Seiler,
Vacuum orbit and spontaneous symmetry breaking in
hyperbolic sigma-models, Nucl. Phys. {\bf B720} (2005) 235;
Erratum, Nucl.~Phys.~{\bf B758} (2006) 330.
%
\bibitem{kupi} A.~J.~Kupiainen, On the $1/n$ expansion,
Commun. Math. Phys. {\bf 73} (1980) 273.
%
\bibitem{SZ}   T.~Spencer and M.~Zirnbauer, Spontaneous symmetry breaking 
of a hyperbolic sigma model in three dimensions, 
Commun.~Math.~Phys.~{\bf 252} (2004) 167 [arXiv:math-ph/0410032]. 
%
\bibitem{NSW} M.~Niedermaier, E.~Seiler and P.~Weisz,
Perturbative and non-perturbative correspondences between compact
noncompact sigma-models, 
Nucl.~Phys.~{\bf B788} (2008) 89 [arXiv:hep-th/0703212].
%
\bibitem{TDlimit} A. Duncan, M. Niedermaier, and P. Weisz,
Noncompact sigma-models -- Large $N$ expansion
and thermodynamic limit, Nucl.~Phys.~{\bf B791} (2008) 193
[arXiv:0706.2929]. 
%
\end{thebibliography}
\end{document}